\documentclass[12pt]{article}

\usepackage{amsmath,amssymb,graphicx} %use drftcite in draftmode
\usepackage{pstricks}
\newcommand{\beq}{\begin{eqnarray}}% can be used as {equation} or  {eqnarray}
\newcommand{\eeq}{\end{eqnarray}}

\newcommand{\centeron}[2]{{\setbox0=\hbox{#1}\setbox1=\hbox{#2}\ifdim
\wd1>\wd0\kern.5\wd1\kern-.5\wd0\fi \copy0
\kern-.5\wd0\kern-.5\wd1\copy1\ifdim\wd0>\wd1
                                    \kern.5\wd0\kern-.5\wd1\fi}}
\newcommand{\ltap}{\>\centeron{\raise.35ex\hbox{$<$}}
                            {\lower.65ex\hbox{$\sim$}}\>}
\newcommand{\gtap}{\>\centeron{\raise.35ex\hbox{$>$}}
                            {\lower.65ex\hbox{$\sim$}}\>}

\newcommand\ZZ{\hbox{\zfont Z\kern-.4emZ}}
\font\zfont = cmss10 %scaled \magstep1

\newcommand{\eref}[1]{eq.\ (\ref{e.#1})}
\newcommand{\erefn}[1]{ (\ref{e.#1})}

\newcommand{\cref}[1]{Chapter \ref{c.#1}}

\def\nn{\nonumber \\}

\def\beq{\begin{equation}}
\def\eeq{\end{equation}}
\newcommand{\ba}{\begin{array}}
\newcommand{\ea}{\end{array}}
\newcommand{\bea}{\begin{eqnarray}}
\newcommand{\eea}{\end{eqnarray} }
\newcommand{\bal}{\begin{align}}
\newcommand{\eal}{\end{align}}
\def\bi{\begin{itemize}}
\def\ei{\end{itemize}}
\def\ben{\begin{enumerate}}
\def\een{\end{enumerate}}
\def\beq{\begin{equation}}
\def\eeq{\end{equation}}
\def\bc{\begin{center}}
\def\ec{\end{center}}
\def\bt{\begin{table}}
\def\et{\end{table}}
\def\btb{\begin{tabular}}
\def\etb{\end{tabular}}

\def\co{{\mathcal O}}

\def\tev{\, {\rm TeV}}

\def\mass2{mass${}^2$}

\def\ra{\rangle}
\def\la{\langle}

\def\pa{\partial}

\def\simlt{\stackrel{<}{{}_\sim}}

\def\eps{\epsilon}

\begin{document}
\begin{titlepage}
%\begin{flushright}
%{\tt hep-ph/yymmnn}
%\end{flushright}

\vskip1.5cm
\begin{center}
{\huge \bf Holographic Unhiggs}
\vspace*{0.1cm}
\end{center}
\vskip0.2cm

\begin{center}
{\bf Adam Falkowski$^{a}$ and Manuel P\'erez-Victoria $^{b}$}

\end{center}
\vskip 8pt

\begin{center}
$^a$ { \it CERN Theory Division, CH-1211 Geneva 23, Switzerland} \\
\vspace*{0.3cm}
$^b$ {\it CAFPE and Departamento de F\'{\i}sica Te\'orica y del Cosmos, \\ 
             Universidad de Granada, E-18071, Spain } 

\vspace*{0.3cm}

{\tt adam.falkowski@cern.ch, mpv@ugr.es}
\end{center}

\vglue 0.3truecm

\begin{abstract}
\vskip 3pt \noindent 

We propose an extra-dimensional approach to the unparticle Higgs (Unhiggs). 
The non-local 4D Unhiggs action is derived as the boundary effective action of a local 5D theory. 
We review the mechanism to generate unparticle spectra in the context of warped soft-wall models. 
Gauge invariance can be simply implemented and unitarity of the longitudinal WW scattering amplitude is particularly transparent. 
Furthermore, the 5D approach uncovers a broader spectrum of phenomenological possibilities. 
The Unhiggs is accompanied by another continuum formed by Kaluza-Klein excitations of the Standard Model gauge fields which,  
in some cases, may take over the leading role in unitarizing WW scattering.

\end{abstract}

\end{titlepage}

%\newpage
\renewcommand{\theequation}{\arabic{section}.\arabic{equation}} 
\renewcommand{\thefootnote}{(\arabic{footnote})}

%%%%%%%%%%%%%%%%%%%%%%%%%%%%%%%%%%%%%%%%%%%%%%%%%%%%%%%%%%5
\section{Introduction}
\setcounter{equation}{0} 
\label{s.i} 
%%%%%%%%%%%%%%%%%%%%%%%%%%%%%%%%%%%%%%%%%%%%%%%%%%%%%%%%%%%5

Various aspects of unparticles \cite{G} have been wildly studied in the past year.
The notion of unparticles refers to a new physics sector that displays a continuous spectrum of excitations 
%(rather than a set of sharply defined particles) 
and is, most often, conformal over a range of scales.   
The main motivation for studying this scenario is that unparticles lead to collider signals that differ dramatically from those encountered in the conventional particle physics.  
At the advent of the LHC, studies of unconventional phenomenology may be justified.
On the other hand, there has been so far little indication that the concept of unparticles may shed some new light on the nagging problems of the Standard Model (SM), in particular on the naturalness problem originating from large quantum corrections to the Higgs boson mass.  
But it is precisely the naturalness problem which is the reason why, in the first place, we expect new physics at the LHC.   
 
Any new physics that addresses the naturalness problem should become manifest not far above the electroweak scale set by the W boson mass.
A part of the expected energy range up to a few hundred GeV has been directly probed by the Tevatron, with the null result.  
Indirect probes, like electroweak precision tests that probe the new physics up to a few TeV, or flavor physics that is even more sensitive to generic flavor and CP violating new physics (up to $10^5$ TeV in some cases \cite{UTfit}), have detected nothing either.
All the new physics scenarios that we are aware of, whether supersymmetry, composite Higgs, or technicolor, are uncomfortable with these negative search results. 
If the new physics addressing the naturalness problem does exist at the LHC reach, it must be stealthy and elusive.   
 
It is conceivable that, if the new physics manifested itself as a blurry unparticle continuum rather than ordinary particles, the sensitivity of the low-energy physics to the scale of new physics might be reduced.  
A hint that this is indeed the case has emerged recently in a model studied in ref.\ \cite{FV} (see also \cite{BGS}). 
In that model the Higgs is a pseudo-Goldstone boson of the usual particle nature, but the vector bosons beyond those of the SM appear as a continuum. 
While electroweak precision tests typically constrain the masses of new vector resonances to be larger than 2-3 TeV, the vector continuum can well start below 1 TeV.  

What if the Higgs itself is an unparticle?  
Such concepts have appeared in the unparticle literature, but most of the previous studies \cite{DEQ} were restricted to a particle Higgs boson mixing with an unparticle sector that is not charged under the SM. 
From the naturalness perspective, this is not very different to mixing the Higgs with an ordinary SM singlet particle.   
A new scenario appears in a model proposed recently by Stancato and Terning \cite{ST} (see also \cite{CDHH}).  
These authors consider a 4D action for an unparticle Higgs (the Unhiggs) that transforms as a doublet under the SM $SU(2)_L$ group, and whose kinetic term in momentum space has the non-local form $(\mu^2 - p^2)^{2 - d}$, with $1< d < 2$. 
Therefore, the Unhiggs tree-level propagator has a branch cut for $p^2$ larger than the mass gap $\mu$ squared, in contrast to the propagator of a scalar particle (corresponding to $d = 1$), which has simply a pole.
It turns out, nevertheless, that this weird object can play the same important role as the ordinary Higgs particle does: it can {\em break the electroweak symmetry} and {\em unitarize WW scattering}.
The former is a simple consequence of  the Unhiggs acquiring a vacuum expectation value (vev).   
Unitarity of WW scattering is less trivial, because covariantization of the Unhiggs action with respect to the SM local symmetry group leads to gauge interactions that are completely different from the SM ones \cite{CMT2}. 
In particular, the Unhiggs has interaction vertices with an arbitrary number of gauge bosons, and the coefficients of these vertices depend on the Unhiggs vev. 
In spite of these unusual features, the exchange of the Unhiggs contributes to the tree-level scattering amplitude of two longitudinally polarized W bosons in a way that renders this amplitude well behaved in the ultraviolet (UV).
The cancellations that lead to the unitary behavior turn out to be quite different than in the SM though:  
both the direct gauge vertex and the Unhiggs exchange amplitude contain non-analytic powers of kinematical invariants, and these two non-analytic contributions cancel against each other at high energies. 

Thus, electroweak symmetry breaking via the Unhiggs is a new concept in physics beyond the SM. 
While at this point it is not clear if it can help with any of the problems plaguing the SM (although ref.\ \cite{ST} suggested that the little hierarchy problem may be reduced), it is interesting to explore this new direction.         

In this paper we propose another description of the Unhiggs. 
Our approach is inspired by the AdS/CFT conjecture, which relates large N strongly coupled theories to weakly coupled 5D gauge theories in a warped background \cite{adscft,APR}.
If such a large N sector gives rise to the SM Higgs sector, there should exist a 5D effective description in the language of Randall-Sundrum (RS) type models \cite{RS1}.  
It was demonstrated before \cite{CMT,FV} that unparticle spectra can be obtained in the so-called {\em soft-wall} version of RS, where the infrared (IR) brane is removed and effectively replaced by an exponentially decaying warp factor \cite{KKSS}.
We adapt this framework to study the Unhiggs.
More precisely, the  Unhiggs scenario refers here to spontaneous breaking of electroweak symmetry triggered by a scalar field whose spectrum of excitations is (partly) continuous.  
Conformal symmetry over a range of scales can be realized if the warped background is approximately AdS in the region near the UV boundary, but this is not a necessary ingredient.    
The focus of this paper is on issues related to longitudinal WW scattering in this scenario. 
%We will investigate various contributions to the amplitude and how they conspire to ensure the unitary behavior of total amplitude.    

In the 5D description, the Higgs sector is represented by a 5D scalar field charged under the electroweak group. The corresponding gauge bosons must then propagate in 5D as well.
This scalar field has a non-trivial potential that forces its vev, thus breaking the electroweak symmetry. 
The 5D set-up can accommodate the SM gauge bosons and fermions of ordinary particle nature, together with Kaluza-Klein (KK) excitations that form a continuum separated by a mass gap from the SM. 
In particular, the oscillations around the scalar vev display a continuous spectrum, 
and they mix with an ordinary particle Higgs living on the UV boundary of the 5D spacetime.  
The 4D low-energy effective description of this set-up corresponds to a non-local Unhiggs action as in ref.\ \cite{ST}.
The connection is most transparent using the boundary effective action approach \cite{BPR}, which amounts to keeping the UV boundary degrees of freedom as the low-energy variables and integrating out the 5D bulk.   
%We demonstrate this  connection at the quantitative level. 

There are several features of our holographic approach that motivate its application to study the Unhiggs.
First of all, the 5D models reveal a broader spectrum of phenomenological possibilities. 
Due to the fact the the electroweak gauge bosons must propagate in the 5D bulk, the spectrum, apart from the Unhiggs, also contains a continuum of vector excitations.   
We argue that, typically, the mass gap for the vector unparticles is smaller than the one for scalars, so that the vector continuum should show up at the LHC along with the Unhiggs.
Moreover, the vector continuum  can also play an important role in the dynamics of electroweak symmetry breaking: in certain regions of the 5D parameter space it gives the dominant contribution to longitudinal WW scattering, much like vector resonances in the Higgsless models \cite{CGPT}.  

It can be argued that the presence of a vector continuum is a generic feature of the Unhiggs scenario, and not just an artifact of the 5D formulation.
The fact that the Unhiggs is charged under the SM electroweak group implies that the hidden sector that gives rise to the Unhiggs has a global $SU(2) \times U(1)$ (or larger) symmetry. 
The electroweak gauge fields couple to the conserved currents of this global symmetry.     
If the hidden sector is strongly interacting, the two-point correlators of these currents have a non-trivial spectral representation that includes resonances and/or a continuum of excitations. 
The latter possibility occurs, in particular, when the hidden sector is conformal over some energy range.  
As a consequence of coupling to the conserved currents, the external electroweak gauge bosons mix with the excitations of the hidden sector (much like the photon mixes with the vector resonances in ordinary  QCD), which effectively leads to interactions between the SM gauge fields and the hidden sector.        
The 5D set-up provides a consistent realization of  this scenario.    

At a more technical level, the advantage of the 5D formulation is that the action is manifestly local and gauge invariant, and the usual methods of perturbative quantum field theory apply. 
There has been some debate in the literature on how to correctly gauge non-local unparticle actions \cite{CMT2,I}. 
In 5D, on the other hand, gauge invariance is trivially implemented by the usual covariant derivative prescription, and it automatically implies that the physical predictions of the effective Unhiggs action are gauge independent.
Simply playing with the parameters of the 5D model we get, almost effortlessly, a complete model of the Unhiggs with all interactions consistently included. 
Moreover, our approach allows us to generate extensions of the original Unhiggs construction; for instance, the Unhiggs self-interactions can easily be made non-local. 
Besides gauge invariance, various consistency issues become more transparent in 5D.
For example, the energy range where the theory is applicable can be precisely defined. 
Furthermore, in the present context we can use with confidence the Goldstone boson equivalence theorem \cite{et} to calculate the longitudinal WW scattering amplitude. 
This allows us to find a very clear picture of how various contributions conspire to make the total amplitude consistent with unitarity. 
In the longer perspective, the 5D approach should facilitate a systematic analysis of electroweak precision observables and collider signals.

Here is the plan. 
In Section \ref{s.u5d}, we introduce our 5D model based on the $SU(3)_C\times SU(2)_L \times SU(2)_R \times U(1)_X$ gauge symmetry and a bifundamental Higgs field in the bulk.
We explain how this framework can be adapted to study unparticles.  
To this end, we review the salient features of the soft-wall models and the necessary conditions to arrive at a continuum spectrum.
We point out the relevance of propagators in mixed position/momentum space to investigate 5D theories with unparticle like spectra.
We explain how integrating out the bulk leads to a 4D effective action of the unparticle type.  
We also present a concrete example of 5D background leading to an unparticle effective action.   
In Section \ref{s.vc}, we turn to vector KK excitations of the SM gauge boson, which are a new indispensable ingredient in the 5D approach. 
In Section \ref{s.ww}, we discuss the longitudinal WW scattering. 
We explain how the Goldstone boson equivalence theorem can be used in the 5D setting.
We show that the holographic Unhiggs (backed by the vector continuum) fulfills the task of unitarizing the WW scattering. 
Our general formalism is illustrated with a specific example, which exactly reproduces the WW scattering amplitude found by Stancato and Terning. 
We conclude in Section \ref{s.c}.

%%%%%%%%%%%%%%%%%%%%%%%%%%%%%%%%%%%%%%%%%%%%%%%%%%%%%%%%%%5
\section{The Unhiggs in 5D}
\setcounter{equation}{0} 
\label{s.u5d} 
%%%%%%%%%%%%%%%%%%%%%%%%%%%%%%%%%%%%%%%%%%%%%%%%%%%%%%%%%%%5

%%%%%%%%%%%%%%%%%%%%%%%%%%%%%%%%%%%%%%%%%%%%%%%%%%%%%%%%%%5
\subsection{Model}
\label{s.sm} 
%%%%%%%%%%%%%%%%%%%%%%%%%%%%%%%%%%%%%%%%%%%%%%%%%%%%%%%%%%%5

We consider a 5D extension of the SM.
The theory propagates in a warped geometry parametrized by the warp factor $a(z)$:   
$ds^2 = a(z)^2 (dx_\mu^2 - dz^2)$, $\mu = 0\dots 3$. 
The 5th coordinate $z$ runs from $R$ (the UV brane) to infinity, but the invariant length $L = \int_R^\infty a(z)$ is assumed finite.
We fix $a(R) = 1$.  
There is no IR brane,  unlike in the standard RS scenario, which opens the door
%clears the way
to realizing unparticles as a continuum of KK excitations \cite{CMT}. We refer to the region of large $z$ as the IR.
The SM electroweak gauge group is embedded in a larger 5D gauge symmetry $SU(2)_L \times SU(2)_R \times U(1)_X$. 
Extending the hypercharge $U(1)_Y$ to $SU(2)_R \times U(1)_X$ is necessary from the phenomenological point of view, in order to avoid excessive contributions to the Peskin-Takeuchi T parameter \cite{ADMS}.
Moreover, we introduce the scalar Higgs field $\Phi$ propagating in the 5D bulk, in the bifundamental representation under $SU(2)_L\times SU(2)_R$.
The bulk lagrangian is 
\beq
\mathcal{L} =
\sqrt{g}\left\{ 
-\frac{1}{2}\text{Tr}\left(L_{MN}L^{MN}+ R_{MN}R^{MN}\right) 
-\frac{1}{4}X_{MN}X^{MN}
+ \frac{1}{4}\text{Tr}\left|D_M\Phi\right|^2 - V(\Phi) 
\right\},  
\eeq
where $D_M \Phi = \pa_M \Phi - i g_{L*}  R^{1/2} L_M \Phi + i g_{R*} R^{1/2} \Phi R_M$.
The gauge couplings in 5D are dimensionful, and we denote them as $R^{1/2} g_{L*,R*,X*}$. 
The SM color group is irrelevant for the present discussion, and it is ignored throughout this paper.  
We are not concerned much with the SM fermions here, but they can be simply included, for example as UV brane-localized chiral fermions 
coupled to the bulk Higgs via brane-localized Yukawa interactions, or as in ref.\ \cite{BGS}. 

In the following we will simplify a bit our task by taking the limit $g_{X*} \to 0$.
This corresponds to taking the zero hypercharge limit in the SM, which  makes the masses of W and Z bosons equal to each other and decouples them from the photon. The latter lives then entirely in the $U(1)_X$ factor.  
The case of general $g_{X*}$ does not introduce any philosophical complications, but makes the relevant formulas lengthier.    
Moreover, we also assume equal couplings for $SU(2)_L$ and $SU(2)_R$: $g_{L*} = g_{R*} \equiv g_*$.
This limit will be enough to illuminate the major points related to WW scattering.

Breaking of the gauge symmetry down to the electromagnetic $U(1)$ is realized via two distinct mechanisms: by boundary conditions and by the Higgs mechanism. 
The UV brane is assumed to respect only the electroweak $SU(2)_L \times U(1)_Y$ subgroup of the full gauge symmetry group.  
%, where the hypercharge is realized as $Y = T_R^3 + X$. 
% $B = s_x R^3 + c_x X$. 
This is done in practice by imposing the Dirichlet boundary conditions at $z = R$ for three generators of $SU(2)_R \times U(1)_X$. 
In general, the unbroken generator is a linear combination of $T_R^3$ and $X$, but in our zero hypercharge limit the entire $SU(2)_R$ is broken on the UV brane, while $U(1)_X$, hosting the photon, is left unbroken.  
On the other hand, the bulk Higgs is endowed with a potential $V(\Phi)$  and the boundary potential $V_{UV}(\Phi)$ that enforce the vev  $\la \Phi \ra  = R^{-1/2} \hat v (z) I_{2 \times 2}$, which spontaneously breaks  $SU(2)_L\times SU(2)_R$ down to the vector $SU(2)$. 
The UV boundary potential $V_{UV}$  allows us to control the boundary conditions for the Higgs field.
For example, if $V_{UV}$ is vanishing the Higgs obeys the Neumann boundary conditions $\pa_z \Phi(R) = 0$, whereas if $V_{UV}$ is very large (for example, the mass term in $V''_{UV}$ is infinite) the boundary conditions become Dirichlet, $\Phi(R) = 0$.
The UV brane lagrangian, in general, also includes the Higgs kinetic term (even if it is set to zero at tree-level it will be induced by loops) and higher-derivative local terms, but these are not important for the following discussion.   

The 5D set-up we have sketched above is dual to a 4D strongly coupled sector with a {\em global} $SU(2)_L\times SU(2)_R \times U(1)_X$ symmetry, whose $SU(2)_L\times U(1)_Y$ subgroup is weakly gauged by fundamental electroweak gauge  fields.
The bulk Higgs field is interpreted as a composite scalar operator emerging from the strongly coupled sector.
The UV boundary condition amounts to mixing the composite operator with a normal particle Higgs field of mass $\sim V''_{UV}$.   
The limit of the Dirichlet boundary condition corresponds to decoupling the fundamental Higgs, leaving only the composite operator.

%%%%%%%%%%%%%%%%%%%%%%%%%%%%%%%%%%%%%%%%%%%%%%%%%%%%%%%%%%5
\subsection{Continuum vs Discretuum}
\label{s.scvd} 
%%%%%%%%%%%%%%%%%%%%%%%%%%%%%%%%%%%%%%%%%%%%%%%%%%%%%%%%%%%5

Coming back to 5D, the Higgs vev should satisfy the equation of motion and boundary conditions 
\beq
\label{e.hve}
\pa_z (a^3 \pa_z \hat v)  - a^5  {\pa V(\hat v) \over \pa  \hat v} = 0,  
\qquad  \pa_z \hat v(R) = R {\pa V_{UV}(\hat v) \over \pa  \hat v},  \qquad   \lim_{z\to \infty} a^3(z) \hat v(z) = 0.   
\eeq 
The Higgs fluctuations around that background are parametrized as $\Phi(x,z) \to [R^{-1/2} \hat v(z) +  \hat h(x,z) ]I$. 
The dynamical scalar field $\hat h$ obeys the equation of motion and boundary conditions
\beq
\label{e.heom} % higgs equation of motion 
\left [  a^{-3} \pa_z (a^3 \pa_z) -   \hat M^2  + p^2 \right ]  \hat h  = 0, 
\qquad   [\pa_z - M_{\rm UV}^2 R] \hat{h}(x,R) = 0, 
\qquad \lim_{z\to \infty} a^3(z) \hat{h}(x,z) = 0,
\eeq  
where $\hat M^2 = a^2 R {\pa^2 V(\hat v) \over \pa \hat v^2}$ and  $ M_{\rm UV}^2 = {\pa^2 V_{UV}(\hat v) \over \pa \hat v^2}$.   

%So far all we have said applies as well to the standard warped theory with a discrete spectrum of resonances, 
%but from this point we are converging to  the continuum case. 
There is a simple method to investigate if an equation of motion leads to a discrete set of KK modes or to a continuous spectrum.
%, and if there is a mass gap separating the continuum from zero.  
The trick consists in rewriting the equation of motion into a form resembling the Schr\"odinger equation by defining the ``wave function" $\hat \Psi = a^{3/2} \hat h$. 
This leads to the equation  
\beq
\left ( - \pa_z^2 + \hat V  \right ) \hat \Psi = p^2 \hat \Psi \, , 
\qquad
\hat V = \hat M^2(z) + {3 a''\over 2 a} +  {3 (a')^2 \over 4 a^2}. 
\eeq 
The Schr\"odinger potential $\hat V(z)$ depends on both the Higgs potential  (via the mass squared term) and the 5D geometry (via the derivatives of the warp factor). 
Now, much as in the textbook quantum mechanics, the Schr\"odinger potential could be an infinite well, in which case the spectrum is discrete, or it could flatten out and allow for a continuum.    
%Since the hard-wall UV brane provides already one infinite wall, 
The kind of spectrum we encounter depends on the asymptotic behavior of $\hat V(z)$ for $z \to \infty$. 
According to this, the spectrum of the Higgs excitations can be classified into three families: 
\ben
\item {\bf Discretuum}, that is a tower of discrete KK modes, for $\hat V(z)|_{z \to \infty} \to \infty$. 
The examples are all hard-wall IR brane scenarios, or the soft wall scenarios where the warp factor decays sufficiently fast,
 $a \sim e^{-(\rho z)^\alpha}$, $\alpha > 1$.  
\item {\bf Unparticles without Mass Gap} for $\hat V(z)|_{z \to \infty} \to 0$. 
The classical example is that with the AdS metric $a(z) = R/z$, which is nothing but a RS2 set-up \cite{RS2}. 
On the 4D side, the corresponding theory is Georgi's original unparticle model \cite{G}.  
\item {\bf Continuum with Mass Gap}, for $\hat V(z)|_{z \to \infty} \to \mu^2 > 0$.  
An example \cite{CMT} is the warp factor $a(z) = e^{- 2\rho z}/z$ with any mass term that vanishes in the IR, leading to a mass gap $\mu = 3 \rho$.   
\een 
There is a cornucopia of model-building based on discretuum, and here we are choosing another direction. 
For unparticles charged under the SM gauge interactions, the lack of a mass gap is at odds with experiment. This brings us to the third case:
in the following we assume that the Schr\"odinger potential asymptotes to a constant $\mu^2>0$ in the IR, 
so that the 5D scalar field $\hat h$ has a continuous spectrum of excitations. This is the holographic Unhiggs scenario.

%%%%%%%%%%%%%%%%%%%%%%%%%%%%%%%%%%%%%%%%%%%%%%%%%%%%%%%%%%5
\subsection{Propagator}
\label{s.sp} 
%%%%%%%%%%%%%%%%%%%%%%%%%%%%%%%%%%%%%%%%%%%%%%%%%%%%%%%%%%%5

Before proceeding, we need to develop some tools to tackle a 5D theory with a continuous spectrum.  
The usual approach to perturbative computations in 5D theories goes via the KK expansion, 
but in the Unhiggs scenario this is not efficient\footnote{Although one could introduce an IR regulator brane to proceed in the usual way, and remove the regulator at the end of the calculation.}. 
Another, less popular approach uses the formalism of 5D propagators defined in a mixed momentum/position representation, which we shortly refer to as p/z propagators.  
In general, the p/z propagator $P(p^2,z,z')$ describes the propagation of the entire KK tower of excitations carrying 4D momentum $p$ between the two points $z$ and $z'$ in the extra dimension.  
It can be calculated by Fourier-transforming the 4D coordinates in the 5D action, and inverting the kinetic term. 
This approach works in the same manner in the unparticle case. An important feature of the propagator is that its analytic structure encodes the entire information about the spectrum. 
In particular, the usual discrete KK resonances correspond to poles in the propagator, while the continuum shows up as a discontinuity of its imaginary part across the real positive $p^2$ axis.

The Unhiggs p/z propagator satisfies the equation   
\beq 
\left [a^{-3}(z) \pa_z \big(a^3(z) \pa_z\big) -   \hat M^2  + p^2 \right ]\hat P(p^2,z,z')  =   a^{-3}(z)  \delta(z - z'),   
\eeq 
and the UV boundary condition $[\pa_z - M_{\rm UV}^2 R]\hat P(p^2,z,z')$. 
It is convenient to formally solve this equation in an arbitrary background.
To this end, we denote the two independent solutions of the equation of motion \erefn{heom} as $\hat K(z,p^2)$, $\hat S(z,p^2)$. The solution
$\hat K(z,p^2)$ is defined such that it asymptotes to $e^{-p z}$ for large Euclidean momenta and is normalized as $\hat K(R,p^2) =  1$.
$\hat S(R,p^2)$ is another independent solution, and it is convenient to choose it such that $\hat S(R,p^2) = 0$, $\hat S'(R,p^2) = 1$.   
In that notation, the Unhiggs propagator for  $z \leq z'$ can be written as  
\beq
\label{e.uhp}
\hat P (p^2,z,z') = 
{\hat K (z,p^2)\hat K(z',p^2)  \over  R \hat \Pi(p^2)} - \hat S(z,p^2) \hat K(z',p^2), 
\qquad
\hat \Pi(p^2) = R^{-1} \hat K'(R,p^2) - M_{\rm UV}^2. 
\eeq 
Note that the propagator naturally splits into a ``boundary" part   
$\hat P^{(B)} = \hat K (z,p^2)\hat K(z',p^2)/R \hat \Pi$, and a
``Dirichlet" part $\hat P^{(D)} = -  \hat S(z,p^2) \hat K(z',p^2)$, which vanishes on the UV boundary.
The boundary-to-boundary propagator is given by $\hat P(p^2,R,R) = 1/R \hat \Pi(p^2)$. 
For reasons that will become clear in a second, we refer to $\hat \Pi(p^2)$ as the {\em kinetic function}.

%%%%%%%%%%%%%%%%%%%%%%%%%%%%%%%%%%%%%%%%%%%%%%%%%%%%%%%%%%5
\subsection{Effective action}
\label{s.sea} 
%%%%%%%%%%%%%%%%%%%%%%%%%%%%%%%%%%%%%%%%%%%%%%%%%%%%%%%%%%%5

There is a straight path from the p/z propagators to the boundary effective action. 
We define the low-energy variable 
%$\bar h(p) =  R^{-1/2} \hat h(p,R)/\hat P(p^2,R,R)$, 
$\bar h(p) =  R^{1/2} \hat h(p,R)$,
where  $\hat h(p,z)$ is the Fourier transform of $\hat h(x,z)$ with respect to 4D coordinates.  
As usual, the quadratic part of the tree-level effective action in momentum space is given by the inverse of the UV-boundary-to-boundary propagator 
\beq
\label{e.seff}
S_{eff} = \frac{1}{2} \int {d^4 p \over (2 \pi)^4} \bar h(-p)  \hat \Pi(p^2)  \bar h(p) + \dots \ . 
\eeq    
The kinetic function $\hat \Pi$ in 5D warped models is non-local, and has in general a non-trivial analytic structure in the complex $p^2$ plane.  
In the soft-wall unparticle scenario, $\hat \Pi(p^2)$ has a branch cut for $p^2 > \mu^2$.  
Thus the effective action is an Unhiggs type action. It is approximately conformal in the UV if the gravitational background is approximately AdS in the vicinity of the UV brane. 
In particular, $\hat \Pi(p^2) \sim (\mu^2 - p^2)^{2- d}$  considered in ref.\ \cite{ST} can be obtained in specific backgrounds, as we will see in a moment. 
Note that only the boundary part of the propagator contributes to the quadratic Unhiggs action.      
The Dirichlet part is not irrelevant however, as it affects the quartic and higher vertices of the gauge fields in the effective action.  
In fact, we will show that the Dirichlet part plays the major role in unitarizing WW scattering. 
%This splitting will prove very important for the sake of comparison between our holographic Unhiggs and  the 4D construction.
  
The formal expressions in eqs.\ \erefn{uhp} and \erefn{seff}  will allow us to  discuss the Unhiggs physics in full generality, without referring to a specific background. 
At the same time, they can also work as a blueprint that we can readily fill in each case when the solutions $\hat K$, $\hat S$ to the equations of motion are explicitly known. 
Below we illustrate our general discussion with a specific example.  
Ideally, one would like to start with a bulk Higgs potential, possibly motivated by string theoretical and/or holographic considerations \cite{BG}, and solve the coupled scalar+gravity equations of motion. 
For the sake of the present discussions, we will be satisfied with an ad-hoc example where the warp factor and the mass terms in the equations of motion are fixed by a simple ansatz.
This is a self-consistent procedure as long as we neglect the gravitational degrees of freedom.

%%%%%%%%%%%%%%%%%%%%%%%%%%%%%%%%%%%%%%%%%%%%%%%%%%%%%%%%%%5
\subsection{Example}
\label{s.se} 
%%%%%%%%%%%%%%%%%%%%%%%%%%%%%%%%%%%%%%%%%%%%%%%%%%%%%%%%%%%5

We now present an example of 5D background that links our holographic approach to the 4D construction of Stancato and Terning. 
We take the warp factor  
$a(z) = {R \over z } e^{-2 \rho (z-R)}$ and assume $\rho R \ll 1$.\footnote{This is just the usual RS hierarchy, which should be justified in the context of a complete scalar-gravity action.} 
Thus, the warp factor interpolates between AdS spacetime for $z \ll 1/\rho$ and the soft-wall IR cut-off at $z \sim 1/\rho$, producing a continuum spectrum with a mass gap $\mu = 3 \rho$.
We choose the bulk Higgs potential
$V(\phi) =  \chi_1(z)(m_\phi^2  - \chi_2(z))\phi^2/2$, where $ m_\phi^2 R^2 = \nu^2 - 4$, and the $z$-dependent functions $\chi_1 = e^{4 \rho (z - R)}$ and $\chi_2 = 9 \rho z/R^2$ originate from some ``dilatonic" vevs.\footnote{This peculiar choice is motivated by the fact that it links directly to the Unhiggs action of ref.\ \cite{ST}. 
We could also use a simpler dilaton background  with $\chi_2 = 0$, in which case we can still solve the equations of motion in terms of the hypergeometric functions, and the boundary Unhiggs action would contain additional non-analytic terms behaving as $\mu/\sqrt{-p^2}$ for momenta above the mass gap.  
For a simple Higgs mass term  corresponding to $\chi_1 = 1$, $\chi_2 = 0$ we were not able to solve the equations of motion analytically. In any of these two cases, the solutions near the UV boundary would be the same as in our example, and thus the effective Unhiggs action would also be the same for $|-p^2| \gg \mu^2$.}  
With this potential the equations of motion can be solved in terms of Bessel functions. 
The vacuum equation \erefn{hve} for the bulk Higgs vev  is solved by 
\beq
\label{e.stvs} % vacuum solution 
\hat v(z) = v_0  a^{-3/2}(z) {z^{1/2} K_\nu(\mu z) \over R^{1/2} K_\nu(\mu R)}, 
\eeq  
and the constant of integration $v_0$ is fixed by the UV boundary condition $ \pa_z \hat v(R) = R V_{UV}'(\hat v)$.
We choose the boundary Higgs potential as in the SM: $V_{UV} = m_{UV}^2 \hat v^2/2 + \lambda_{UV} \hat v^4/4$. 
Then the UV boundary conditions are solved by  
\beq
\label{e.stv} 
\lambda_{UV} v_0^2  = {2 - \nu \over R^2} - m_{UV}^2  + {\mu \over  R} \left ( 1 - { K_{1 - \nu}(\mu R) \over K_{\nu}(\mu R)}\right ), 
\eeq   
or by $v_0 = 0$. 
If the right-hand side of \eref{stv} is positive, then $v_0^2 > 0$ and the electroweak symmetry is broken.    
%If  which determines $v_0$ in terms of the parameters of the 5D model.
Note that $m_{UV}^2 < 0$ is not a necessary condition to trigger electroweak symmetry breaking.  
 
Next, the bulk Higgs mass term is $\hat M^2 (z) = {\nu^2 - 4 \over z^2} - {9 \rho \over z}$, which leads to the following solutions to the Higgs equation of motion \erefn{heom}: 
\bea
\label{e.ste} % ST example
\hat S (z,p^2) &=&  R a^{-3/2}(z) (z/R)^{1/2}
\nn && 
\cdot \left [K_\nu(\sqrt{\mu^2 - p^2} R) I_\nu(\sqrt{\mu^2 - p^2}z) - I_\nu(\sqrt{\mu^2 - p^2} R) K_\nu(\sqrt{\mu^2 - p^2}z)
\right ],
\nn
\hat K (z,p^2) &=& a^{-3/2}(z) (z/R)^{1/2}{K_\nu(\sqrt{\mu^2 - p^2}z) \over K_\nu(\sqrt{\mu^2 - p^2} R) } .   
\eea 
Inserting these solutions into \eref{uhp} we obtain the Unhiggs propagator. 
The kinetic function is given by  
\beq 
\label{e.stk} % ST kinetic
\hat \Pi(p^2) = 
 {\mu  K_{1 - \nu}(\mu R) \over R K_{\nu}(\mu R)} 
-  {\sqrt{\mu^2 - p^2} K_{1 - \nu}(\sqrt{\mu^2 - p^2} R) \over R K_{\nu}(\sqrt{\mu^2 - p^2} R) } 
- M_{uh}^2, 
% - C(\nu) R^{2 \nu - 1} m^{2 \nu} 
\eeq 
where $M_{uh}^2 = 2 \lambda_{UV} v_0^2$.  
It is non-analytic in $p^2$ and picks up an imaginary for $p^2 > \mu^2$, 
so the boundary effective action \erefn{seff} is of the unparticle type with mass gap $\mu$.
Apart from the mass gap, the Unhiggs of \eref{stk} is characterized by another mass scale $M_{uh}$ that is an analogue of a mass term, in the sense that $\hat \Pi(0) \sim -M_{uh}^{2}$.
This role of $M_{uh}$ is most transparent for $|p|$ much smaller than the mass gap: 
in that case \eref{stk} reduces to a normal particle action, $\hat \Pi(p^2) \approx Z p^2 - M_{uh}^2$ with 
$Z = [K_{1 - \nu}(\mu R) K_{1 + \nu}(\mu R)/K_{\nu}(\mu R)^2  - 1]/2$.
The spectral density of the propagator is positive definite (no ghosts)
and, as long as $M_{uh}^2 \geq  0$, it  has support in the positive real axis (no tachyons).      
It is clear that we need $M_{uh}^2 \geq 0$ to avoid instabilities, but otherwise we do not find any consistency constraints; in particular $M_{uh}$ can be much larger than the mass gap.     

When the parameter $\nu$ is in the range $0 < \nu < 1$, the quadratic boundary effective action at low energies reduces to the Unhiggs free action of ref.\ \cite{ST}.  
Indeed, for $|p| R \ll 1$ the kinetic function can be approximated as    
\beq
\label{e.stk1}
\hat \Pi(p^2)  \approx   R^{2 \nu - 2} C(\nu) \left [ \mu^{2 \nu} - (\mu^2 - p^2)^\nu -  m^{2 \nu} \right ],  % + \co 
\eeq    
where $C(\nu) = 2^{1 - 2\nu} \Gamma(1 -\nu)/\Gamma(\nu)$, and  $m^{2 \nu}  = M_{uh}^2 R^{2 - 2 \nu} /C(\nu)$.
After rescaling of $\bar h$ by $R^{-\nu + 1/2}/C(\nu)^{1/2}$ this is precisely the Unhiggs of ref.\ \cite{ST} with a scaling dimension $d = 2 - \nu$,\footnote{More precisely, the AdS/CFT interpretation in terms of a dual operator of dimension $d = 2 - \nu$ is appropriate once the mass $m$ has been tuned to small values, so that the Unhiggs propagator is approximately $ \hat \Pi^{-1} \sim (-p^2)^{-\nu}$. 
If $m$ is very large, one would rather expand the propagator as $\hat \Pi^{-1} \sim - 1/m^{-2 \nu} + (-p^2)^{2 \nu}/m^{-4 \nu}$ and conclude that we deal with a dual operator of dimension $d = 2 + \nu$ plus a contact term. 
This is related to the usual ambiguity in the AdS/CFT dictionary \cite{KW}.} and  the UV scale $\Lambda$ identified with $R^{-1}$.
 
Since eq. \erefn{stk} is valid for arbitrary $\nu \geq 0$, we do not have to restrict our analysis to $0 < \nu < 1$.  
In a sense, \eref{stk} provides a ``continuation" of the Unhiggs action beyond the open interval $1 < d < 2$.    
In particular, even though the limit $\nu \to 0$ of \eref{stk1} is singular, the same limit in \eref{stk} is perfectly smooth.  
For $|p| R \ll 1$ the $\nu = 0$ case can be approximated as     
\beq
\hat \Pi(p^2)  \approx  
{1 \over R^2 \log (e^\gamma \sqrt{\mu^2 -p^2} R/2)} - {1 \over R^2 \log (e^\gamma\mu R/2)}
  -  M_{uh}^2.
\eeq 
Similarly, one should note that the limit $\nu = 1$ in the full holographic expression  differs from the usual $d = 1$ particle by the presence of non-analytic logarithmic corrections: 
\beq
\hat \Pi(p^2)  \approx  (\mu^2 - p^2) \log (e^\gamma \sqrt{\mu^2 - p^2} R/2)  - \mu^2 \log (e^{\gamma}\mu R/2) 
- M_{uh}^2 . 
\eeq 
For $1 < \nu < 2 $ the kinetic function in the same regime is approximated by 
\beq
\label{e.stk3} 
\hat \Pi(p^2)  \approx  
{1 \over 2\nu} p^2 +  R^{2 \nu - 2} C(\nu) \left [ \mu^{2 \nu} - (\mu^2 - p^2)^\nu -  m^{2 \nu} \right ].
\eeq   
This time, the leading term is the usual particle kinetic term, while the non-analytic terms are suppressed by $(|p| R)^{2 \nu - 2}$ and therefore they are subleading.   
This illustrates our previous comments that in the 5D approach consistency comes almost for free.      
The 5D model ``knows" that the Unhiggs action with $d < 1$ cannot be consistent, and thus it ``remembers" to include a healthy analytic kinetic term in the effective action that dominates over the non-analytic one.\footnote{ 
In order to make the non-analytic terms dominant one would have to add a ghost-like kinetic term in the UV brane Higgs action, with a coefficient tuned to cancel the $p^2$ term  in \eref{stk3}.}
As a result, the case with $\nu > 1$ is, in practice, not much different from an ordinary particle Higgs. 
%: regarding the physical properties and the fine-tuning likewise.
Going to $\nu > 2$, higher integer powers of $p^2$ appear in the expansion, which are also dominant over the non-analytic terms but suppressed with respect to the $p^2$ term.

Finally, let us discuss the fine-tuning of the mass term in the Unhiggs action.  
From naturalness arguments one would expect $m_{UV} \sim v_0 \sim R^{-1}$, which leads to $M_{uh} \sim m \sim R^{-1}$.
For phenomenological reasons (and also for perturbativity of the WW scattering, as we will see later), $m$ is required to be not much larger than the electroweak scale. 
This can be obtained by fine-tuning the brane mass term $m_{UV}^2$. 
The degree of this fine-tuning depends on the parameter $\nu$ and becomes less severe as we approach $\nu = 0$.     
In the limit $\nu = 0$, one needs very mild fine-tuning, $M_{uh}^2 R^2 \sim 1/\log (\mu R)$.
% to make the ``kinetic term" dominate over the ``mass term" for $p^2$ above the mass gap.   
However, one should not fire champagne corks yet. 
As we will discuss later in Section 4, fitting the W boson mass to the observed value requires fine-tuning that actually becomes worse in the limit $\nu\to 0$.  

%In addition, $\hat \Pi$ contains other $(\mu^2 - p^2)^{2 \nu}$ suppressed by  $|p| R \ll 1$  

%%%%%%%%%%%%%%%%%%%%%%%%%%%%%%%%%%%%%%%%%%%%%%%%%%%%%%%%%%5
\section{Vector Continuum}
\setcounter{equation}{0} 
\label{s.vc} 
%%%%%%%%%%%%%%%%%%%%%%%%%%%%%%%%%%%%%%%%%%%%%%%%%%%%%%%%%%%5 

In the 5D setting there is a crucial difference compared to the 4D approach of ref.\ \cite{ST}.  
The Unhiggs born from a 5D bulk must always be accompanied by 5D gauge bosons, which give rise not only to the SM gauge bosons, but also to either a discretuum or a continuum of vector excitations. 
Which of these two possibilities arises is determined by the same method we discussed before in the context of the Unhiggs, that is by the Schr\"odinger potential for the gauge bosons. 
The potential can be found by transforming the gauge boson equations of motion into the Schr\"odinger form, and it is different for the vector  and axial combinations,  $\sqrt{2} V_M = (L_M + R_M)$, $\sqrt{2} A_M = (L_M - R_M)$, since only the latter couples to the bulk Higgs vev. 
We easily find \cite{FV}  
\beq
V_{\rm vector}  = {a''\over 2 a} - {(a')^2 \over 4 a^2}, 
\qquad 
V_{\rm axial} = M^2 + {a''\over 2 a} - {(a')^2 \over 4 a^2},
\eeq
where $M^2(z) = g_{*}^2 a^2 \hat v^2/2$.  The potential for $U(1)_X$ excitations is the same as for the vector ones.  
%Note that the UV boundary conditions mix the vector and the axial components, so that the mass eigenstates are in general mixtures of vector and axial excitations. 

The vector Schr\"odinger potential depends only on the warp factor.
% and, unlike the axial one, does not see the electroweak symmetry breaking vev. 
From that we conclude that in the 5D Unhiggs scenario the warp factor should decay {\em precisely} as $e^{- 2 \rho z}$ (up to power-law terms in $z$) in the far IR.
Indeed if it decayed faster than the exponential, it would affect the Unhiggs Schr\"odinger potential too and prevent the existence of the Unhiggs continuum (except in the extremely fine-tuned situation where a negative growing mass function $\hat M^2$ precisely cancels the growing terms generated by the warp factor in the Schr\"odinger potential).  If, on the other hand, it decayed slower than the exponential, then the photon would not be separated by a mass gap from the continuum.  
Thus, {\em in 5D the vector continuum with a mass gap $\rho$ is typically present when the Unhiggs continuum is present}.  
Recall that the Unhiggs mass gap  is $\mu = \sqrt{(3 \rho)^2  + \hat M^2(\infty)}$. 
If $\hat M^2(z)$ vanishes in IR, the Unhiggs continuum starts at $3\rho$, that is three times above the vector continuum. 
The situation where the Unhiggs mass gap is smaller than the vector mass gap is not impossible, but it requires some engineering.
Namely, the mass squared should asymptote to a negative value satisfying $-9 \rho^2 < \hat M^2(\infty) < - 8\rho^2$.       
Clearly, the more generic situation is the one where the vector continuum is encountered first. 
This is a very interesting consequence of the holographic Unhiggs scenario.  
  
Since we have to deal with the continuum spectrum of gauge bosons, it is convenient to introduce the corresponding p/z propagators. 
They  can be computed by the same token as the Unhiggs propagator, that is by inverting the quadratic terms in the 5D action. 
The additional complication is that we deal with vector and axial excitations that follow different equations of motion 
\bea
\label{e.veom}
\left [ a^{-1} \pa_z (a \pa_z) + p^2  \right ] f_{V} &=& 0,
\\ 
\label{e.aeom}
\left [ a^{-1} \pa_z (a \pa_z) + p^2 - M^2(z)\right ] f_{A} &=& 0,
\eea  
and the two are mixed by the UV boundary conditions
\beq
\label{e.vabc}
\pa_z \left  [ f_V(R) + f_A(R) \right ]  = 0, 
\qquad 
f_V(R) -   f_A(R)   = 0 .
\eeq 
%where,  for simplicity,  we assumed no UV boundary terms for the gauge bosons.   
Hence, the propagator is a $2\times2$ matrix in the $(V,A)$ space.     
As before, it is convenient to  write the propagators in terms of the formal solutions to the equation of motion.  
We define $K_{M}(z,p^2)$ as the axial equation of motion \erefn{aeom} that is regular in the IR (it decays exponentially for large Euclidean momenta) and normalized as $K_M(R,p^2)= 1$. Similarly, $K_{0}(z,p^2)$ is the IR regular solution to the vector equation \erefn{aeom}.
$S_{0}(z,p^2)$ ($S_{M}(z,p^2)$) is another independent solution to the vector (axial) equation of motion with the UV boundary conditions $S_{\cdot}(R,p^2)= 0$, $S_{\cdot}'(R,p^2)= 1$. 
%In the following we will need only the VV propagator. 
We use the Feynman gauge, in which the propagators are diagonal in the Lorentz indices, $P_{\mu\nu} = \eta_{\mu\nu} P$. 
After some algebra we find, for $z \leq  z'$,      
\bea 
\label{e.vvp}
P_{VV}(p^2,z,z') &=& {K_0(z,p^2)  K_0(z',p^2) \over \Pi_V(p^2)}  -  S_0(z,p^2)  K_0(z',p^2),
\nn
P_{AA}(p^2,z,z') &=& {K_M(z,p^2)  K_M(z',p^2) \over \Pi_V(p^2)}  -  S_M(z,p^2)  K_M(z',p^2),
\nn 
P_{VA}(p^2,z,z') &=& {K_0(z,p^2)  K_M(z',p^2) \over \Pi_V(p^2)};  \qquad \Pi_V(p^2) = K_M'(R,p^2) + K_0'(R,p^2). 
\eea  
Again, the propagators naturally split into boundary and Dirichlet parts (the $VA$ propagator has only boundary part because the vectors and axials only mix through the UV boundary conditions).
Each propagator has a pole for $p^2 = m_W^2$, defined by the lowest zero of $\Pi_V(p^2)$. 
It also has other singularities on the positive $p^2$ axis corresponding to the KK excitations.   
From the previous discussion we know that it has a branch cut (signaling the presence of a continuum)  whenever the Unhiggs propagator has a branch cut.  

With the propagators at hand we construct the boundary effective action. 
We define the low energy $SU(2)$ gauge bosons as 
%$\bar L_{\mu}(p) = L^{-1/2} L_\mu(p,R)/2 P_{VV}(p^2,R,R)$.
$\bar L_{\mu}(p) = L^{1/2} L_\mu(p,R)$.
Integrating out the bulk at tree-level, we get the quadratic effective action
\beq
\label{e.vba}
S_{eff} = \frac{1}{2} \int{d^4 p \over (2 \pi)^4} \bar L_{\mu}(-p) \left[ {1 \over 2 L} \Pi_V(p^2) \right ] \bar L_{\mu}(p) + \dots \ .   
\eeq
In general, this effective action includes the effects of integrating out the vector continuum: the kinetic function picks up an imaginary part for $p^2$ larger than the mass gap $\rho^2$. If $\rho$ is much larger then the electroweak scale (as we discussed, this not generic in 5D if the Unhiggs mass gap is at weak scale) we can neglect the effects of the continuum. 
In practice, this is obtained by projecting all gauge propagators to the zero-mode propagator $P(p^2,z,z') \to 1/(2 L p^2)$.   
   
The dots in \eref{vba} stand for an infinite number of interaction terms, which follow from the interaction vertices in the 5D theory.  
They can be obtained in the following way: propagate the fields from the boundary to the bulk, insert the 5D vertices, connect them with propagators, and integrate over all possible vertex positions.
For instance, two axial gauge bosons interact with one 5D Higgs field via $(g_{*}^2/2) R^{1/2} a^3 \hat v \hat h A_{M}^2$. 
This leads to the following interaction term between two gauge bosons and one Unhiggs in the effective theory: 
\beq 
S_{eff}^{hV^2} = c \int {d^4 q_1 \over (2 \pi)^{4}} {d^4 q_2 \over (2 \pi)^{4}} \bar  L_{\mu}(q_1) \bar  L_{\mu} (q_2) \bar h(-q_1 - q_2). 
\eeq    
The coefficient of this vertex can be computed from the integral 
\beq
c = {g_*^2 \over 4 L} \int_R^\infty dz a^3(z) \hat v(z) 
{P_{VA}(q_1^2,R,z) \over  P_{VA}(q_1^2,R,R)} {P_{VA}(q_2^2,R,z) \over P_{VA}(q_2^2,R,R)}
{\hat P(p^2,R,z) \over \hat P(p^2,R,R)}. 
\eeq   
In principle, given a 5D background we can solve the equations of motion, find the propagators and compute the overlap integrals to find the interaction terms in the 4D boundary effective theory.  
Performing this exercise in the example defined in \eref{ste} we would find that the $\bar h \bar L^2$ vertex matches the one in ref.\ \cite{ST} if we project the vector propagators inside the integral to the zero-mode propagators, that is if we neglect the effects of the vector continuum. Similarly, the non-local quartic vertex in the boundary effective action contains in this example the modified four-W vertex of ref.\ \cite{ST}.

%%%%%%%%%%%%%%%%%%%%%%%%%%%%%%%%%%%%%%%%%%%%%%%%%%%%%%%%%%5
\section{WW scattering}
\setcounter{equation}{0} 
\label{s.ww} 
%%%%%%%%%%%%%%%%%%%%%%%%%%%%%%%%%%%%%%%%%%%%%%%%%%%%%%%%%%%5

We move to the focus point of this paper, which is the mechanism by which the holographic Unhiggs ensures unitarity of WW scattering.
One could perform this calculation using the Unhiggs boundary effective action, but we find it more convenient to take advantage of the local formulation and compute the amplitude directly in 5D.       
To tackle this question we will use the equivalence theorem, which says that the scattering of longitudinally polarized vector bosons is equivalent, at leading order in $E^2/v^2$, to the scattering of the Goldstone bosons eaten by the gauge bosons. 
In the SM, the eaten Goldstone bosons are the components of the Higgs doublet corresponding to $SU(2)$ rotations of the vacuum state.  
In 5D theories things are a bit more complicated \cite{FPR}, because the Goldstone bosons are distributed between the bulk Higgs field $\Phi$ and the 5th component $V_z^a,A_z^a$ of the vector and axial gauge fields. 
We first explain how to extract the couplings  of the eaten Goldstone boson in 5D theories, 
and then compute the longitudinal WW scattering amplitude. 
We use our general results to compute the amplitude in the example introduced earlier in Section \ref{s.se}.

%%%%%%%%%%%%%%%%%%%%%%%%%%%%%%%%%%%%%%%%%%%%%%%%%%%%%%%%%%5
\subsection{Goldstones}
\label{s.sg} 
%%%%%%%%%%%%%%%%%%%%%%%%%%%%%%%%%%%%%%%%%%%%%%%%%%%%%%%%%%%5

The Goldstone components in the bulk Higgs $\Phi$ are made explicit in the non-linear parametrization    
\beq
\Phi(x,z) =  \left (R^{-1/2}\hat v(z) + \hat h(x,z) \right ) e^{i G^a \sigma_a} . 
\eeq 
In fact, $G^a$, $V_z^a$,$A_z^a$ host the entire tower of Goldstones, that are eaten by the KK excitations of the gauge fields.\footnote{%
They also host physical pseudo-scalars which are not relevant for WW scattering, and we can ignore them here.} 
In order to extract the Goldstones eaten by the W and Z we have to perform a KK expansion, much as for the gauge bosons.       

To find the profile of an eaten Goldstone, the first step is to find the profile of the corresponding gauge boson. 
In the zero hypercharge approximation that we are assuming here, the W and Z bosons have the same profile and the same mass. 
They are embedded in both the vector and the axial component of the 5D gauge fields,   
$V_\mu^a (x,z) \to   W_{\mu}^a(x) f_{V}(z)$, 
$A_\mu^a (x,z) \to   W_{\mu}^a(x) f_{A}(z)$   
The profiles satisfy the equations of motion \erefn{veom}, \erefn{aeom} with $p^2 = m_W^2$   
and the UV boundary conditions \erefn{vabc}.  
The ``IR boundary conditions" in the soft-wall case amount to choosing the IR regular solutions (the ones denoted by $K_{\cdot}(z,p^2)$) of the differential equations \erefn{veom} and \erefn{aeom}.  
Hence, the profiles can be written as 
\beq
f_V(z) = \alpha K_0(z,m_W^2), \qquad f_A(z) = \alpha K_M(z,m_W^2),
\eeq 
where $\alpha$ is fixed by the normalization condition $\int_R^\infty a (f_V^2 + f_A^2) = 1$. 
The relation between $m_W$ and the 5D parameters is determined by the UV boundary conditions, which
lead to the quantization condition $\Pi_V(m_W^2) \equiv K_M'(R,m_W^2)  + K_0'(R,m_W^2) = 0$.  
    
Now, given the profiles of the W and Z gauge bosons, the corresponding Goldstones $G^a(x)$ are embedded into the 5D fields as   
\bea
\label{e.gp} % goldstone profile 
V_z^a(x,z)  &\to&  m_W^{-1} \pa_z f_{V}(z) G^a(x), 
\nn
A_z^a(x,z)  &\to&  m_W^{-1} \pa_z f_{A}(z) G^a(x),
\nn
G^a(x,z) &\to & m_W^{-1} \sqrt{R/2} g_{*} f_{A}(z) G^a(x). 
\eea
This embedding ensures 1) that the kinetic terms for the Goldstones combine with the W,Z mass terms into  
${1 \over 2} \left ( \pa_\mu G^a - m_W W_{\mu}^a \right )^2$, 2) that $G^a(x)$ does not mix with the higher KK excitations, and  
3) that there is no mass term for $G^a(x)$. 
These three properties define $G^a(x)$ as the Goldstone boson eaten by $W_\mu^a$.

%%%%%%%%%%%%%%%%%%%%%%%%%%%%%%%%%%%%%%%%%%%%%%%%%%%%%%%%%%5
\subsection{Amplitude}
\label{s.sa} 
%%%%%%%%%%%%%%%%%%%%%%%%%%%%%%%%%%%%%%%%%%%%%%%%%%%%%%%%%%%5

Having obtained the Goldstone profiles, we can find the vertices relevant for WW scattering. 
This is done by inserting the Goldstone profiles \erefn{gp} into the following 5D interactions terms:   
\bi 
\item {\bf Goldstone self-coupling}
\beq
{a^3 \hat v^2 \over 6 R}  \left ( \pa_\mu G^a \pa_\mu G^b G^a G^b - \pa_\mu G^a \pa_\mu G^a G^b G^b   \right ),   
\eeq   
\item {\bf Higgs vertex}
\beq 
{a^3 \hat v \over R^{1/2}}  \hat h \pa_\mu G^a \pa_\mu G^a,   
\eeq 
\item {\bf Vector vertex}
\beq
-  {a R^{1/2} g_* \over \sqrt{2}}  
\eps^{abc}  \left (\pa_\mu V_z^a V_z^b   + \pa_\mu A_z^a A_z^b  + a^2 \hat v^2  \pa_\mu G^a G^b \right ) V_\mu^c.   
\eeq 
\ei

We now have everything we need to compute the longitudinal WW scattering amplitude. 
Consider the Goldstone scattering process $G^a G^b \to G^c G^d$.  
The $SU(2)$ structure of this amplitude can be factored out,  
\beq
M_{ab\to cd} = \delta_{ab} \delta_{cd} M(s,t,u) + \delta_{ac} \delta_{bd} M(t,u,s)  + \delta_{ad} \delta_{bc} M(u,s,t), 
\eeq
so that the amplitude (in the zero hypercharge limit) is unambiguously described by one function $M(s,t,u)$ of the Mandelstam kinematical variables. 
The amplitude gets contributions from the Goldstone quartic self-interactions, the Unhiggs exchange and the vector boson exchange:  
$M(s,t,u)= M_G + M_h + M_V$.  
Propagation of the Unhiggs and the vector bosons between the vertices is captured by the corresponding p/z propagators, 
and we have to integrate over all possible positions of each vertex in the 5th dimension. 
We find 
\bea
\label{e.wwa}
M_{G}(s) &=&  {g_{*}^2 R \over 2 m_W^4} \left (s - {4 m_W^2 \over 3} \right)  \int_{R}^\infty  a M^2(z) f_A^4,  
\nn 
M_{h}(s) &=& -  {g_{*}^2 R \over 2  m_W^4}  (s-2 m_W^2)^2  \int_{R}^\infty dz dz'  \phi_h (z)  \phi_h (z') P_h(s,z,z'),  
\nn
M_{V}(s,t,u) &=& - {g_{*}^2 R \over 2  m_W^4 }  
\left ( (s - u) \int_{R}^\infty dz dz'  \phi_g (z) \phi_g (z' ) P_{VV}(t,z,z') 
\right . \nn  && \left .
+ (s - t) \int_{R}^\infty dz dz'  \phi_g (z) \phi_g (z' ) P_{VV}(u,z,z') 
 \right ), 
\eea 
where 
\beq
\phi_h (z) =  a^2 M(z) (f_A)^2,  
\qquad 
\phi_g (z) =
a \left [ (\pa_z f_V)^2 + (\pa_z f_A)^2  +  M^2(z) (f_A)^2 \right ].
\nonumber
\eeq 
This is the complete tree-level scattering amplitude given in terms of 5D input parameters.
Once the 5D background is chosen, we can always  use \eref{wwa} to calculate (perhaps numerically) the longitudinal WW scattering amplitude. In the following we will analytically extract some general physical properties of the amplitude. 

%%%%%%%%%%%%%%%%%%%%%%%%%%%%%%%%%%%%%%%%%%%%%%%%%%%%%%%%%%5
\subsection{Discussion}
\label{s.d} 
%%%%%%%%%%%%%%%%%%%%%%%%%%%%%%%%%%%%%%%%%%%%%%%%%%%%%%%%%%%5

Let us first take a look at the low energy limit of these amplitudes.
More precisely, we consider the scattering energy $E$ above the W mass but smaller than the Unhiggs ($\mu$) and the vector ($\rho$) mass gaps.  
In that regime, the amplitude should grow quadratically with energy.    
In fact, the low energy theorems for WW scattering \cite{CGG} and the custodial symmetry fix the low energy behavior of the amplitude to  be
$M(s,t,u) \approx s g_L^2/4 m_W^2$, up to subleading corrections suppressed by the mass gap. 
One obvious contribution of this type is that of the Goldstone self-coupling amplitude 
\beq
M_G(s) \approx s {g_{L}^2 L \over 2 m_W^4} \int_{R}^\infty  a M^2(z) f_A^4.
\eeq    
This is not, in general, equal to $s g_L^2/4 m_W^2$, so that there must be other $\co(s)$ contributions to satisfy the low energy theorems.   
One can also see that the Unhiggs exchange is not $\co(s)$ but rather of order $s^2/m_W^2\mu^2$ at low energies. 
However, a not-so-obvious $\co(s)$ contribution is included in the vector exchange amplitude $M_V$.   
At small $p^2$, the solutions to the vector equation of motion \erefn{veom} are approximately momentum independent, $K_0 (z,p^2) \approx 1$, $S_0 (z,p^2) \approx \int_R^z a^{-1}$, so that the Dirichlet part of the vector propagator is  $P_{VV}^{(D)} \approx  \int_R^z a^{-1}$. 
This leads to an $O(s)$ contribution given by 
\beq
M_V^{(D)}(s) \approx s {3 g_{*}^2 R \over  m_W^4}  \int_{R}^\infty dz' \phi_g (z')  \int_{R}^{z'} dz \phi_g (z) \int_R^z a^{-1},
\qquad s \ll \rho^2. 
\eeq  
This contribution can be thought of as that of an effective four-W vertex obtained after integrating out the vector continuum. 
Using the equations of motion and integrating by parts it can proven that $M_G$ and $M_V^{(D)}$ at low energies combine to the total amplitude of the form required by the low energy theorems.
   
There is one important class of 5D models where $M_G$ saturates the low energy theorem and the vector contribution is subleading. 
This is the case when electroweak breaking in the 5D bulk is perturbative.
That is to say, the mass term $M^2(z)$ in the axial equations of motion can be treated  as a perturbation, and, in consequence, the IR regular solution at small $p^2$ can be expanded as 
$K_M(z,p^2) = 1 + \int_R^z a^{-1} \int_R^{z'} a [p^2 - M^2(z'')] + \co(p^4,M^4)$. 
This leads to the approximate expressions for the W profile and mass
\beq
f_A(z) \approx {1 \over \sqrt{2 L}}, \qquad m_W^2 \approx {1 \over 2 L} \int_{R}^\infty a(z) M^2(z)  ,
\eeq 
which trivially give $M_G(s) \approx s g_L^2/4 m_W^2 + \co(m_W^0)$.
The vector contribution is expected to become relevant whenever $M^2$ is in some sense large, for example if it grows in the IR such that    
$\int a M^2$ diverges. 

Let us now turn to the question how the $\co(s)$ contribution is canceled above the Unhiggs and the vector mass gap.
Recall that in the SM the Higgs exchange diagram reads  $M_h \approx -   {g_L^2  \over 4 m_W^2} {s^2 \over s -  m_h^2}$.  
For $s > m_h^2$ this cuts off the $O(s)$ growth of the total amplitude: in short, the SM Higgs unitarizes WW scattering.  
In the present case, unitarization is performed together by the Unhiggs and the vector KK excitations.   
It is fairly easy to prove that, for asymptotically UV momenta, the  $O(s)$ term cancel in the total scattering amplitude.   
This follows from the definition of the p/z propagators and the fact that at large $p^2$ the Schr\"odinger equation is approximately solved by the simple exponentials $e^{\pm i p z}$.  
This leads to an asymptotic large $p^2$ formula for the propagators,  
\beq 
\label{e.pass}
P_h (p^2,z,z') \approx {1 \over p^2} a^{-3}(z)  \delta(z-z'), 
\qquad 
P_{VV} (p^2,z,z') \approx {1 \over p^2} a^{-1}(z) \delta(z-z'). 
\eeq
Plugging this into \eref{wwa} we see that the Unhiggs exchange $M_h$ exactly cancels the  $\co(s)$ terms from the Goldstone self-coupling amplitude $M_G$.  
On the other hand, the vector exchange amplitude $M_V$ at high energies (unlike at low energies) does not contribute any $\co(s)$ term.
These conclusions are valid for an arbitrary 5D background. 
%Thus, the total amplitude $M$ asymptotes to a constant. 
%If the constant is small enough, unitarity is preserved. 
%That question depends on  how fast the propagators reach the asymptotic form \erefn{pass}.
%The behaviour of the propagator right above the mass gap is model dependent, but we can study it in explicit examples. 

We can divide the 5D scenario into two distinct classes, depending whether the Unhiggs continuum or the vector continuum gives the dominant contribution to unitarize WW scattering. 
Let us define the quantity
\beq
\alpha_{H/L} =  {2  L \over m_W^2} \int_{R}^\infty  a M^2 f_A^4,  
\eeq  
which is the leading Goldstone self-coupling amplitude in the 5D model divided by the low energy amplitude $s g_L^2/4 m_W^2$.
We can distinguish two situations 
\bi 
\item $\alpha_{H/L} \simlt 1$. 
In that case it is the scalar continuum that provides the dominant contribution to tame the $\co(s)$ growth of the low energy WW scattering amplitude.
As we discussed,  $\alpha_{H/L} \approx 1$ whenever electroweak breaking in 5D can be treated perturbatively.
This is a generalization of the SM Higgs mechanism to the continuum case, for which ref.\ \cite{ST} coined the name the {\em Unhiggs  scenario}.  
\item $\alpha_{H/L} \ll 1$.  
Then it is rather the vector continuum who is responsible for unitarizing WW scattering.  
Since in this case the Unhiggs does not couple significantly to the electroweak gauge bosons, this is clearly a version of Higgsless theories \cite{CGPT}, where the KK modes of electroweak gauge bosons are a continuum rather than a set of discrete resonances.  
Consequently, this case should be referred to as the {\em Unhiggsless  scenario} ;-)  
\ei 

Of course, the fact that $\co(s)$ terms in the longitudinal WW scattering always cancel {\em asymptotically} does not yet guarantee perturbative unitarity of the 5D theory. 
For example, the continuum may kick in too late to save unitarity (much like the SM  Higgs much heavier $1 \tev$ is not consistent with unitarity). 
Less trivially, the UV limits of the Unhiggs and the vector propagators may contain non-analytic powers of momenta that, although subleading with respect to the $\delta(z-z')/p^2$,  may lead to contributions to WW scattering that grow in UV.
Cancellation of these non-analytic terms should be checked separately.   
The above issues are clearly model dependent and must be studied for each specific 5D background. 

Finally, one should remember that tree-level perturbativity in 5D theories is always lost at some energy scale. 
This follows from the fact that the gauge and the Yukawa couplings are dimensionful.
For example, a tree-level amplitude for a process involving gauge fields and sources localized near $z = z_s$ in the 5th dimension involves the gauge coupling $g_*^2 R$ and the propagator $P(p^2,z_s,z_s)$. 
From dimensional analysis and from the UV behavior of the propagator \erefn{pass} we find that  
the amplitude for such a process should be proportional to $g_{*}^2 E R a^{-1}(z_s)$. 
This becomes non-perturbative at the energy scale $E = \Lambda (z_s) \sim 16 \pi^2 g_{*}^{-2} a(z_s) R^{-1}$ and, in general, 
$\Lambda(z_s)$ can be smaller then the UV scale $R^{-1}$.       
Therefore, 5D gauge theories must be treated as effective theories with a cut-off $\Lambda(z_s)$ that is position dependent \cite{RSc}.
In particular, in the far IR of soft-wall models the cut-off scale becomes smaller than the mass gap. 
This does not mean, however, that the theory does not make sense, but rather that there is a limited set of observables that can be meaningfully computed withing the 5D framework. 
For example, computing the effective four-fermion operator induced by the exchange of the vector continuum for fermions localized in far IR would be meaningless.    
On the other hand, for processes localized near the UV brane, $z_s \sim R$, the amplitude remains calculable up to energies parametrically larger than the mass gap.  

%%%%%%%%%%%%%%%%%%%%%%%%%%%%%%%%%%%%%%%%%%%%%%%%%%%%%%%%%%5
\subsection{Example}
\label{s.se2} 
%%%%%%%%%%%%%%%%%%%%%%%%%%%%%%%%%%%%%%%%%%%%%%%%%%%%%%%%%%%5

We close this general discussion with a sample computation % that links our holographic Unhiggs to the 4D Unhiggs of ref.\ \cite{ST}. 
based on the example introduced earlier in Section \ref{s.se}. 
In that example we were able to solve the equations of motion in terms of the Bessel functions. 
For $0 < \nu < 1$ and $|p^2| R^2 \ll 1$, the boundary effective action turned out to of the Unhiggs type, see \eref{stk1}.
%$\sim \bar h^2[\mu^{2\nu}(\mu^2 - p^2)^{\nu} -  m^{2 \nu}]$,    
The vacuum equation for the bulk Higgs vev $\hat v(z)$ could be solved by 
$\hat v \sim  v_0 a^{-3/2}z ^{1/2} K_\nu(\mu z)$, see \eref{stvs}.
Note that $\int a M^2$ is finite, because the Bessel function $K_\nu$ is exponentially damped at large $z$. 
In the following we assume that $v_0$ is small enough that we can treat $M^2(z)$ perturbatively.
This implies that we can express the electroweak scale $v$ in terms of the 5D parameters as   
\beq 
v^2 \approx R^{-1} \int_{R}^\infty a(z)^3 \hat v^2(z)
= {v_0^2 \over 2} \left [ {K_{1 + \nu}(\mu R) K_{1 - \nu}(\mu R) \over K_{\nu}(\mu R)^2} - 1 \right ],   
\eeq 
which leads to $v \sim v_0$ for $\nu > 1$ and $v \sim v_0 (\mu R)^{\nu - 1}$ for $0 < \nu < 1$.
From \eref{stv}, the natural value of $v_0$ is  $\sim R^{-1}$, which is unacceptable. 
For $\nu > 1$ this leads to the usual hierarchy problem: 
$m_{UV}^2$ in the brane lagrangian has to be fine-tuned in order to lower the W mass from the natural UV scale $R^{-1}$ down to the electroweak scale.   For $\nu < 1$ the electroweak scale $v$ is enhanced with respect to $v_0$, so that we need even more fine-tuning than in the SM! 
So, our conclusion is that approaching $\nu \to 0$, that is for the Unhiggs dimension approaching $d= 2$, we lose rather than gain on naturalness. 
On the other hand, once we tune $v \sim \mu$, the mass term in the Unhiggs action given by $m^{\nu} \sim v_0 R^{1 - \nu}$ becomes suppressed with respect to the mass gap in the limit $\nu \to 0$.         
 
In the perturbative case $\alpha_{H/L} \approx 1$ and we can ignore the contribution of the vector continuum to the WW scattering amplitude.
It turns out that in this example we are able to analytically perform the integrals over $z$ in the Unhiggs exchange amplitude.    
It is illuminating to split that amplitude into the ``boundary"  and the ``Dirichlet" parts, corresponding to the analogous splitting of the Unhiggs propagator. 
Keeping the leading term in $m_W^2$, the result is given by    
\begin{align}
M_G(s) \approx & \;   s {g_L^2 \over 4  m_W^2}, 
\nn 
M_h^{(B)}(s) \approx & \; -  {g_L^2 \over 4  m_W^2}  {v_0^2 \over v^2} {\left [\hat \Pi(s) + M_{uh}^2 \right]^2 \over \hat \Pi(s)}, 
\nn 
M_h^{(D)}(s) \approx& \;  -  s {g_L^2 \over 4 m_W^2}
+  {g_L^2 \over 4 m_W^2}{v_0^2 \over  v^2} \left [\hat \Pi(s) + M_{uh}^2 \right] .
\end{align}
Note that the $\co(s)$ term  from the Goldstone self-coupling amplitude $M_G$ is exactly canceled by the Dirichlet part of the Unhiggs exchange amplitude, in agreement with our general arguments.
There are other ``dangerous" terms as the kinetic function $\hat \Pi(s)$ grows at large $s$, in particular, 
$M_h^{(B)}(s)$ may grow with a non-analytic power of $s$. 
It is clear, however, that the ``bad" UV behavior cancels out in the total amplitude:
as soon as $\hat \Pi(s) > M_{uh}^2$, the growth of $M_h^{(B)}$ is canceled by the second term in   $M_h^{(D)}$.
Let us see this explicitly for $0 < \nu < 1$  and $|s| R^2 \ll 1$, in which case the total amplitude can be approximatate as 
\beq
\label{e.stwwa}
M(s,t,u) \approx  {g_L^2 \over 4  m_W^2} {\mu^{2-2\nu}  \over \nu}
 \left [(\mu^2 - s)^\nu - \mu^{2\nu} \right ] 
\left \{ 
{(\mu^2 - s)^{\nu} - \mu^{2\nu}    
\over  (\mu^2 - s)^\nu - \mu^{2 \nu} + m^{2 \nu} } 
-   1 
\right \}
+ \co(m_W^0) + \co(R^{2 \nu}).
\eeq     
The first term in the curly brackets comes from the boundary part, the second from the Dirichlet part. 
Each of them separately grows as $(-s)^\nu$ above the mass gap, 
but the growing terms perfectly cancel once $s > m^2$.
These two terms are exactly the non-analytic contributions found in ref.\ \cite{ST}: the boundary part corresponds to the Unhiggs exchange amplitude in the 4D model, 
while the Dirichlet part corresponds to the part coming from the modified  four-W vertex in the 4D model.
This demonstrates that the 4D construction of ref.\ \cite{ST} can be recovered as a special case of the holographic approach.

There are a few more things worth noting about the amplitude \erefn{stwwa}. 
Firstly, for $s \ll \mu^2$ we recover the usual SM amplitude with the Higgs mass squared replaced by $m^{2 \nu} \mu^{2-2\nu}/\nu$. 
Thus, if the mass gap is much larger than the electroweak scale, the Unhiggs behaves like a particle Higgs, for all practical purposes.      
Next, the cancellation of the non-analytic growing terms would never occur in the limit $m \to \infty$. This is the case of the Dirichlet boundary conditions on the UV brane, which can be interpreted as decoupling the fundamental Higgs living on the boundary and leaving only the composite Higgs from the bulk.  
Finally, for $m \to 0$ the total amplitude vanishes at the leading order in $m_W^2$, due to our choice of a quadratic bulk potential (which is dual to local scalar self-interactions). 
The reason is that, in this limit, apart from the Unhiggs continuum the propagator includes a zero mode Higgs with a flat profile. Then the Unhiggs continuum, being orthogonal to the Higgs zero mode, is also orthogonal to the SM W and Z bosons. Thus, the triple vertex of the Higgs continuum with two W boson vanishes, and only the zero mode Higgs contributes to WW scattering. The cancellation of the growing pieces for finite $m$ can be understood from the $m\to 0$ limit, and the fact that $m$ is a dimensionful parameter.

%%%%%%%%%%%%%%%%%%%%%%%%%%%%%%%%%%%%%%%%%%%%%%%%%%%%%%%%%%%%%%%%%%%%%%%%%%
\section{Conclusions}
\setcounter{equation}{0} 
\label{s.c} 
%%%%%%%%%%%%%%%%%%%%%%%%%%%%%%%%%%%%%%%%%%%%%%%%%%%%%%%%%%%%%%%%%%%%%%%%%% 

In summary, we have discussed the 5D approach to the Unhiggs.
We reviewed the mechanism to generate unparticle-type spectra in the context of soft-wall warped models.   
We developed quite powerful techniques to study this class of theories: 
the mixed momentum/position space propagators defined in arbitrary soft-wall backgrounds, and the 5D version of the Goldstone boson equivalence theorem.  
We also made explicit the link to the 4D Unhiggs construction via the boundary effective action. 

The advantage of the 5D approach is that it allows us to construct, almost effortlessly, unparticle scenarios that are automatically gauge invariant and fully consistent. 
The cancellations leading to a unitary WW scattering amplitude are particularly transparent. 
Furthermore, the 5D parameter space uncovers a broader spectrum of phenomenological possibilities. 
The Unhiggs appears alongside with another continuum consisting of KK excitations of the SM gauge fields, 
and the mass gap of this vector continuum is typically smaller than that of the Unhiggs.    
In some cases, the vector continuum can take over the leading role in unitarizing WW scattering, leading to a new class of models that we called Unhiggsless.
In this paper we did not give a 5D background leading to the Unhiggless case, but this class certainly deserves a dedicated study.   

The fact that the Unhiggs can unitarize WW scattering does not put it on equal footing with the SM Higgs yet. 
One should not forget that the latter also provides loop contributions to electroweak precision observables that, as long as the SM Higgs is not much heavier than 115 GeV, fit very well into precision measurements performed at LEP and the Tevatron.
The real challenge is to find an alternative scenario that would effectively do {\em both}:
unitarize WW scattering {\em and} naturally fit the electroweak precision observables.  
This issue will be addressed in a separate publication.

\section*{Acknowledgments}

We thank Roberto Contino for useful discussions.  
A.F. is partially supported by the European Community Contract MRTN-CT-2004-503369 for the years 2004--2008. 
M.P.V. is supported in part by MEC project FPA2006-05294 and Junta de Andaluc\'{\i}a projects FQM 101, FQM 00437 and FQM 03048.

%&&&&&&&&&&&&&&&&&&&&&&&&&&&&&&&&&&&66

\end{document}